\documentclass[english,eqsecnum,prd,aps,nofootinbib,superscriptaddress,longbibliography,tightenlines,12pt]{revtex4-2}
\usepackage{amsmath}
\usepackage{amssymb}
\usepackage{caption,subcaption}
\usepackage[T1]{fontenc}
\usepackage[latin9]{inputenc}
\usepackage{ulem}
\usepackage[dvipsnames]{xcolor}
\usepackage{array}
\usepackage{booktabs}

\usepackage{babel}
\usepackage{graphicx}
\usepackage[unicode=true,pdfusetitle,
bookmarks=true,bookmarksnumbered=false,bookmarksopen=false,
breaklinks=false,pdfborder={0 0 0},pdfborderstyle={},backref=false,colorlinks=true,allcolors=blue]
{hyperref}

\begin{document}

        \title{Black hole interior quantization: a minimal uncertainty approach}

        \author{Pasquale Bosso}
        \email{pbosso@unisa.it}
        \affiliation{Dipartimento di Ingegneria Industriale, Universit\`a degli Studi di Salerno, Via Giovanni Paolo II, 132 I-84084 Fisciano (SA), Italy}
        \affiliation{INFN, Sezione di Napoli, Gruppo collegato di Salerno, Via Giovanni Paolo II, 132 I-84084 Fisciano (SA), Italy}
        
        \author{Octavio Obreg\'on}
        \email{octavio@fisica.ugto.mx}
        \affiliation{Departamento de F\'isica, Divisi\'on de Ciencias e Ingenier\'ias, Universidad de Guanajuato
                Loma del Bosque 103, Le\'on 37150, Guanajuato, M\'exico}
        
        \author{Saeed Rastgoo}
        \email{srastgoo@ualberta.ca}
        \affiliation{Department of Physics, University of Alberta, Edmonton, Alberta T6G 2G1, Canada}
        \affiliation{Department of Mathematical and Statistical Sciences, University of Alberta, Edmonton, Alberta T6G 2G1, Canada}
        \affiliation{Theoretical Physics Institute, University of Alberta, Edmonton, Alberta T6G 2G1, Canada}

        \author{Wilfredo Yupanqui}
        \email{w.yupanquicarpio@ugto.mx}
        \affiliation{Departamento de F\'isica, Divisi\'on de Ciencias e Ingenier\'ias, Universidad de Guanajuato
                Loma del Bosque 103, Le\'on 37150, Guanajuato, M\'exico}
        
        \begin{abstract}
                In a previous work we studied the interior of the Schwarzschild black
                hole implementing an effective minimal length, by applying a modification to the Poisson brackets of the theory.
                In this work we perform a proper quantization of such a system.
                Specifically, we quantize the interior of the Schwarzschild black hole in two ways: once by using the standard quantum theory, and once by following a minimal uncertainty approach. Then, we compare the obtained results from the two approaches.
                We show
                that, as expected, the wave function in the standard approach diverges in the region
                where classical singularity is located and the expectation value of
                the Kretschmann scalar also blows up on this state in that region.
                On the other hand, by following a minimal uncertainty quantization approach, we obtain 5 new and important results as follows. 
                1) All the interior states remain well-defined and square-integrable. 2) The expectation value of the Kretschmann scalar on the states
                remains finite over the whole interior region, particularly where used to be the classical singularity, therefore signaling
                the resolution of the black hole singularity. 3) A new quantum number is found which plays a crucial role in determining the convergence of the norm of states, as well as the convergence and finiteness of the expectation value of the Kretschmann scalar. 4) A minimum for the radius of the (2-spheres in the) black holes is found 5) By demanding square-integrability of states in the whole interior region, an exact relation between the Barbero-Immirzi parameter and the minimal uncertainty scale is found.

        \end{abstract}
        \maketitle

        \section{Introduction}
        
        Several approaches to quantum gravity predict the existence of a minimal
        measurable length.  
        Among these are candidate theories such as loop quantum gravity (LQG) \cite{Thiemann:2007pyv,Rovelli:2014ssa}, polymer quantization \cite{Ashtekar:2002sn,Corichi:2007tf,Tecotl:2015cya,Morales-Tecotl:2016ijb} which can be considered as motivated by the latter (see below), and string theory \cite{Polchinski:1998rq,Polchinski:1998rr}.
        In particular, in LQG, the gravitational
        sector is written in terms of the holonomies of a certain connection,
        the Ashtekar-Barbero connection, which is an $su(2)$-valued one-form.
        This way, gravity is cast into a Lie gauge theory. In polymer quantization, one rather mimics this method by using LQG techniques when dealing with finite dimensional systems or the matter sector, particularly scalar fields. In another word, works with operators that can be thought of as the exponential of the canonical variables. Hence, LQG and polymer quantization are very closely related.
        On the other hand, a minimal length is implemented in phenomenological models of quantum gravity by modifying the ordinary uncertainty principle of quantum mechanics to accommodate deformations at high energies, thus resulting in a generalized uncertainty principle (GUP) \cite{Maggiore:1993rv,Kempf:1994su,Scardigli:2003kr,Mignemi:2011wh,Hossenfelder:2012jw,Pramanik:2013zy,Bosso:2020aqm,Wagner:2021bqz,Barca:2021epy,Bosso:2021koi,HerkenhoffGomes:2022bnh,Segreto:2022clx,Bosso:2023aht}.
        This is often achieved by modifying the commutation relation between ordinarily canonically conjugated variables.
        Equivalently, it can be considered as an alternate quantization procedure.
        
        One of the main places where the introduction
        of a minimal length has profound consequences is the interior of
        the black holes. Such modifications are usually expected to lead to the
        resolution of singularities and other novel results such as bounce
        from a black hole to a white hole. In LQG, there have been extensive
        studies of both the interior \cite{Modesto:2005zm,Ashtekar:2005qt,Bohmer:2007wi,Corichi:2015xia,Chiou:2008nm,Morales-Tecotl:2018ugi,Blanchette:2020kkk}
        and the full spacetime \cite{Kelly:2020uwj,Gambini:2020nsf,Ashtekar:2018cay,Gambini:2009ie,Modesto:2008im,Alonso-Bardaji:2022ear,Bodendorfer:2019jay}
        of the Schwarzschild black hole, as well as lower dimensional black
        holes mimicking the Schwarzschild full spacetime \cite{Gambini:2009vp,Corichi:2015vsa,Corichi:2016nkp}.
        Since the theory is written in terms of the holonomy of the connection
        and not the connection itself (loosely, the exponential of the connection),
        such a holonomy can only generate finite transformations in the conjugate
        variable, which is the (densitized) triad. Hence, the geometry and geometrical
        operators, such as the area and volume operator only admit discrete
        eigenvalues similar to angular momentum in the usual quantum mechanics.
        This discreteness leads to the resolution of singularities in black holes. In finite dimensional cases, such as cosmological models (and also the Schwarzschild interior),
        this method essentially reduces to the polymer quantization, which again
        leads to resolution of cosmological singularities. Since the interior
        of the Schwarzschild black hole is isometric to the Kantowski-Sachs
        cosmological model, this polymer technique can also be applied, which usually not only leads to the resolution of
        the singularity, but also a bounce from the black hole to a white
        hole, although not all the models lead to a bounce.
        
        Black holes have also been studied 
        using GUP-inspired techniques, although
        to our knowledge not as extensively as in LQG.
        Applying semiclassical techniques
        to the interior of the Schwarzschild written in terms of Ashtekar-Barbero
        connection shows that the singularity can be resolved with such techniques
        \cite{Bosso:2020ztk,Blanchette:2021vid}. However, as a comparative
        study of the interior of the Schwarzschild black hole shows, although
        some GUP approaches lead to similar qualitative results as LQG, not
        all of them actually resolve the singularity \cite{Rastgoo:2022mks}.
        
        In this work, our aim is to compare the standard quantization of the
        Schwarzschild interior with a particular GUP quantization, and especially, to derive the corresponding wave functions of the model in
        each case and computing the spectrum of the Kretschmann scalar operator on such states. The paper is organized as follows: In Sec. \ref{sec:Schw-int}, we present a brief overview of the classical treatment of the interior of the schwarzschild black hole in terms of Ashtekar-Barbero variables. In Sec. \ref{Sec:Quantize-Standard}, we quantize this classical model using the standard quantum mechanical approach, i.e., using the standard commutation relation between the canonical variables. Sec. \ref{Sec:Quantize-GUP} is the heart of the paper. It is dedicated to the quantization of the model using the minimal uncertainty approach, where we derive the wave function of the interior and show that it, as well as the the spectrum of the Kretschmann scalar for this state, is nowhere divergent in the interior. Finally in Sec. \ref{Sec:Conclude} we summarize our results and conclude.
        
\section{The interior of the Schwarzschild black hole\label{sec:Schw-int}}
        
As is well-known, the metric of the interior of the Schwarzschild
        black hole can be obtained by exchanging $t\leftrightarrow r$ in
        the Schwarzschild coordinates, yielding
        \begin{equation}
                ds^{2}=-\left(\frac{2GM}{t}-1\right)^{-1}dt^{2}+\left(\frac{2GM}{t}-1\right)dr^{2}+t^{2}\left(d\theta^{2}+\sin^{2}\theta d\phi^{2}\right).\label{eq:sch-inter}
        \end{equation}
        This is a special case of a Kantowski-Sachs cosmological spacetime
        \cite{Collins:1977fg} 
        \begin{equation}
                ds_{KS}^{2}=-N(T)^{2}dT^{2}+g_{rr}(T)dr^{2}+g_{\theta\theta}(T)d\theta^{2}+g_{\phi\phi}(T)d\phi^{2},\label{eq:K-S-gener}
        \end{equation}
        in which $T$ is a generic time coordinate corresponding to a foliation
        associated to the lapse $N(T)$. Such a metric represents a spacetime with spatial
        homogeneous but anisotropic foliations. As is expected, this represents
        a minisuperspace model, i.e., one with finite number of configuration
        variables. Next, we would like to write the model in Ashtekar-Barbero
        variables. These are the Ashtekar-Barbero connection $A_{a}^{i}=\Gamma_{a}^{i}+\gamma K_{a}^{i}$
        and the desitized triads $\tilde{E}_{i}^{a}$, which are the configuration
        and momenta variables, respectively. Here $\Gamma_{a}^{i},\,K_{a}^{i}$
        are the (Hodge dual of) the spin connection, and the extrinsic curvature
        of the foliations, respectively (see below). The free parameter $\gamma$
        is called the Barbero-Immirzi parameter. Indices $a,\,b,\,c,\,d$
        are the spatial indices and $i,\,j,\,k,\,l$ correspond to the internal
        $su(2)$ gauge freedom of the theory. One can first derive the (undesitized) triads using their definition $q^{ab}=\delta^{ij}E_{i}^{a}E_{j}^{b}$,
        where $q^{ab}$ is the inverse of the spatial part of the above metric.
        Densitized triads $\tilde{E}_{i}^{a}$ are then simply computed
        by multiplying $E_{i}^{a}$ by the square root of the determinant
        of $q_{ab}$ denoted by $\sqrt{q}$, i.e., $\tilde{E}_{i}^{a}=\sqrt{q}E_{i}^{a}$.
        Using triads and the spatial metric, one can then compute $\Gamma_{a}^{i}$
        and $K_{a}^{i}$ using
        \begin{align}
                \Gamma_{a}^{i}= & \frac{1}{2}\epsilon^{i}{}_{jk}E^{bk}\left(\partial_{b}E_{a}^{j}-\partial_{a}E_{b}^{j}+E^{cj}E_{al}\partial_{b}E_{c}^{l}\right),\\
                K_{a}^{i}= & \frac{1}{2}E^{bi}\mathcal{L}_{T}q_{ab}.
        \end{align}
        Here $\mathcal{L}_{T}$ is the Lie derivative with respect to the time direction vector $T^a=Nn^a+N^a$ where $n^a$ is the unit timelike normal vector to the spatial foliations and $N^a$ is the shift vector. Replacing these in the aforementioned formula for $A_{a}^{i}$, one obtains
        the Ashtekar-Barbero connection. It turns out that in this case, there
        are only two independent components to $A_{a}^{i}$, traditionally
        called $b(T),\,c(T)$, with their conjugate momenta being $p_{b}(T)$
        and $p_{c}(T)$. More explicitly, one can write
        \begin{align}
                A_{a}^{i}\tau_{i}dx^{a}= & \frac{c}{L_{0}}\tau_{3}dr+b\tau_{2}d\theta-b\tau_{1}\sin\theta d\phi+\tau_{3}\cos\theta d\phi,\label{eq:A-AB}\\
                \tilde{E}_{i}^{a}\tau_{i}\partial_{a}= & p_{c}\tau_{3}\sin\theta\partial_{r}+\frac{p_{b}}{L_{0}}\tau_{2}\sin\theta\partial_{\theta}-\frac{p_{b}}{L_{0}}\tau_{1}\partial_{\phi},\label{eq:E-AB}
        \end{align}
        where $\tau_{i}$ are the basis of $su(2)$ algebra. Given the topology
        $\mathbb{R}\times\mathbb{S}^{2}$ of the spatial sector of the model,
        in order to prevent the integral involved in computation of the symplectic
        structure in $\mathbb{R}$, one needs to restrict the radial part
        to an auxiliary length $L_{0}$. This restricts the volume of the
        integration to $V_{0}=a_{0}L_{0}$, where $a_{0}$ is the area of
        the 2-sphere $\mathbb{S}^{2}$ in $L_{0}\times\mathbb{S}^{2}$ \cite{Ashtekar:2005qt}.
        This is possible due to homogeneity of the model since one can take
        the limit of $L_{0}\to\infty$ at the end. Clearly, none of the physical
        results should depend on $L_{0},\,a_{0}$ or their rescalings. 
        
        Substituting these into the full Hamiltonian of gravity written in
        Ashtekar-Barbero connection variables, one obtains the symmetry reduced
        Hamiltonian constraint adapted to this model as \cite{Ashtekar:2005qt}
        \begin{equation}
                H=-\frac{N\mathrm{sgn}(p_{c})}{2G\gamma^{2}}\left[2bc\sqrt{|p_{c}|}+\left(b^{2}+\gamma^{2}\right)\frac{p_{b}}{\sqrt{|p_{c}|}}\right],\label{eq:H-class-N}
        \end{equation}
        while the diffeomorphism constraint vanishes identically due to homogeneity. Furthermore, the Poisson brackets between the
        conjugate variables turns out to be 
        \begin{align}
                \{c,p_{c}\}= & 2G\gamma, & \{b,p_{b}\}= & G\gamma.\label{eq:classic-PBs-bc}
        \end{align}
        The Schwarzschild interior metric in these variables is expressed
        as 
        \begin{equation}
                ds^{2}=-N(T)^{2}dT^{2}+\frac{p_{b}^{2}}{L_{0}^{2}|p_{c}|}dr^{2}+|p_{c}|\left(d\theta^{2}+\sin^{2}\theta d\phi^{2}\right),\label{eq:Black.Hole.Metric.In.Ashtekar.Varbl}
        \end{equation}
        and its classical singularity resides at $p_{b}\to0,p_{c}\to0$. 
        
\section{Standard Quantization of the black hole interior\label{Sec:Quantize-Standard}}

\subsection{States}        
        As a first attempt, we would like to quantize this model in the standard fashion using
        the so called Schr\"{o}dinger representation, and find the corresponding
        wave function. This will allows us to contrast this result with the
        GUP wave function that we will derive in the next section.
        
        In classical theory, one can choose any lapse function $N$ and the
        physical content of the theory will not change based on this choice.
        There is a particular choice of the lapse function that makes deriving
        analytical solutions to the equation of motion easier, by rendering
        the differential equations in $c,\,p_{c}$ independent of the differential
        equation in $b,\,p_{b}$. This choice of the lapse reads 
        \begin{equation}
                N=\text{sgn}(p_{c})\sqrt{|p_{c}|}.\label{eq:Lapse-function}
        \end{equation}
        Using this, the Hamiltonian constraint \eqref{eq:H-class-N} becomes
        \begin{equation}
                H=-\frac{1}{2G\gamma^{2}}\left[\left(b^{2}+\gamma^{2}\right)p_{b}+2cbp_{c}\right].\label{eq:Quantum-Hamiltonian}
        \end{equation}
        In our previous work \cite{Bosso:2020ztk} we considered
        the above Hamiltonian and applied the GUP modifications to the classical
        Poisson algebra of the theory to derive an effective description of the interior based on GUP method. In this work we actually quantize this same Hamiltonian using
        both the standard quantization techniques and the GUP methods. 
        
        For the standard non-GUP quantization, we apply the Dirac prescription
        to the classical algebra \eqref{eq:classic-PBs-bc} to obtain the commutation
        algebra of the canonical variables as 
        \begin{align}
                [b,p_{b}]= & iG\gamma, & [c,p_{c}]= & 2iG\gamma, & [b,p_{c}]= & [c,p_{b}]=0,\label{eq:Usual_Commut_Relation}
        \end{align}
        where we have set $\hbar=1$. As a consequence, in a representation
        where $b$ and $c$ act multiplicatively, $p_{b}$ and $p_{c}$ are
        given by differential operators
        \begin{align}
                p_{b}= & -iG\gamma\frac{\partial}{\partial b}, & p_{c}= & -2iG\gamma\frac{\partial}{\partial c}.
        \end{align}
        The representation of the Hamiltonian constraint acting on a generic state yields a  differential equation, which, by considering a generic
        factor ordering parametrized by the dimensionless quantity $a$, can
        be written as \cite{PhysRevD.28.2960} 
        \begin{equation}
                \left[(b^{2}+\gamma^{2})^{1-a}\frac{\partial}{\partial b}\left(b^{2}+\gamma^{2}\right)^{a}+4bc^{1-a}\frac{\partial}{\partial c}c^{a}\right]\psi(b,c)=0.\label{Diff.Eq.Conn.Repr}
        \end{equation}
        Note that $a=1/2$ corresponds to symmetrical ordering, $a=1$ yields
        the case where the configuration variables reside on the rights of
        momenta, and $a=0$ corresponds to the case where the momenta reside
        on the rights of the configuration variables.

        Applying variable separation to \eqref{Diff.Eq.Conn.Repr}, $\psi(b,c)=S(b)R(c)$, one obtains two independent differential equations, one for $b$
        \begin{equation}
                \left(b^2+\gamma^2\right)\frac{dS(b)}{db}+b\left(2a-m\right)S(b)=0,
        \end{equation}
        and one for $c$
        \begin{equation}
                4c\frac{dR(c)}{dc}+\left(4a+m\right)R(c)=0,
        \end{equation}
        where $m$ is a dimensionless parameter. We will see that this parameter is a new quantum number of our model which determines whether the singularity is resolved, and also plays a crucial role in non-divergence of the wave function of the interior of black hole.
        The respective solutions to these two differential equations are
        \begin{align}
                S_m(b)=&C_{1_m}\left(b^2+\gamma^2\right)^{\frac{m}{2}-a},\label{eq:Usual_b_solution}\\
                R_m(c)=&C_{2_m}c^{-\frac{m}{4}-a},\label{eq:Usual_c_solution}
        \end{align}
        where $C_{1_m}$ and $C_{2_m}$ are two integration constants.
        Then, the solution to \eqref{Diff.Eq.Conn.Repr} is given by 
        \begin{equation}
                \psi_{m}(b,c) = A_{m}\left(b^{2}+\gamma^{2}\right)^{\frac{m}{2}-a}c^{-\frac{m}{4}-a},
                \label{Wave.Fun.Sol}
        \end{equation}
        where $A_{m}=C_{1_m}C_{2_m}$.
        Although we have found the wave function of the interior
        of a Schwarzschild black hole in standard quantum mechanics as a function of the
        configuration variables $b,c$, more direct physical information about the black hole
        can be obtained from the momentum variables $p_{b}$ and $p_{c}$, as the metric
        \eqref{eq:Black.Hole.Metric.In.Ashtekar.Varbl} is indeed written
        in terms of such variables. It is therefore convenient to change the representation
        by performing a Fourier transform 
        \begin{equation}
                \bar{\psi}_{m}(p_{b},p_{c})=\int_{-\infty}^{\infty}\int_{-\infty}^{\infty}e^{-ip_{b}b}e^{-ip_{c}c}\ \psi_{m}(b,c)\ db\ dc.\label{Four.Trans}
        \end{equation}
        Replacing \eqref{Wave.Fun.Sol} in \eqref{Four.Trans}, we obtain 
        \begin{multline}
                \bar{\psi}_{m}(p_{b},p_{c})
                = A_{m}\frac{\sqrt{\pi}2^{\frac{1}{2}(3-2a+m)}\gamma^{\frac{1}{2}(1-2a+m)}\Gamma\left(1-a-\frac{m}{4}\right)\sin\left[\pi\left(a+\frac{m}{4}\right)\right]}{e^{i\frac{\pi}{8}(m+4a)}\Gamma\left(a-\frac{m}{2}\right)}\\
                \times \left[\text{sgn}(p_{c})+1\right] \left|p_{c}\right|^{-1+a+\frac{m}{4}}\left|p_{b}\right|^{-\frac{1}{2}(1-2a+m)}K_{-\frac{1}{2}(1-2a+m)}(\gamma\left|p_{b}\right|),\label{eq:Wav.Sol.Mod.Bess.Two.Without.GUP}
        \end{multline}
        where $K_{\nu}(x)$ is the modified Bessel function of the second kind and we have used the property that $K_{-\nu}(z)=K_{\nu}(z)$.
        Such a solution is obtained under the conditions: $(p_{b},p_{c})\in\mathbb{R}$, $\gamma^{2}>0$, $0<a+\frac{m}{4}<1$ and $2a-m>0$.
        We then see that no solution is admissible for $a=0$, \emph{i.e.}, for an ordering with the coordinates on the left of the momenta. Furthermore, we can observe a number of features.
        First, the wave function vanishes for $p_{c}<0$. Moreover, since the exponent $-1+a+\frac{m}{4}$ of $p_{c}$ is always negative
        by the conditions above, the value $p_{c}=0$ is not in the domain of the wave function.
        Thus, we can safely consider the case $p_{c}>0$
        and write 
        \begin{multline}
                \bar{\psi}_{m}(p_{b},p_{c})=A_{m}\frac{\sqrt{\pi}2^{\frac{1}{2}(5-2a+m)}\gamma^{\frac{1}{2}(1-2a+m)}\Gamma\left(1-a-\frac{m}{4}\right)\sin\left[\pi\left(a+\frac{m}{4}\right)\right]}{e^{i\frac{\pi}{8}(m+4a)}\Gamma\left(a-\frac{m}{2}\right)}\times\\
                p_{c}^{-1+a+\frac{m}{4}}\left|p_{b}\right|^{-\frac{1}{2}(1-2a+m)}K_{-\frac{1}{2}(1-2a+m)}(\gamma\left|p_{b}\right|).\label{eq.Wave_func_pc_pb_Usual}
        \end{multline}
        Moreover, for large values of $|p_{b}|$, the wave function vanishes since the Bessel function $K_{\nu}$ vanishes for large values of its argument, according to
        \begin{equation}
                K_{-\frac{1}{2}(1-2a+m)}(\gamma\left|p_{b}\right|)\sim \sqrt{\frac{\pi}{2\gamma|p_b|}}\ e^{-\gamma|p_b|}.
        \end{equation}
        On the other hand, when $p_{b}\to0$, we find 
        \begin{align}
                K_{\frac{2a-m-1}{2}}(|p_{b}|) & \sim\left(\frac{1}{2}\right)^{\frac{3-2a+m}{2}}\Gamma\left(\frac{2a-m-1}{2}\right)\ |p_{b}|^{\frac{1-2a+m}{2}}, & \text{when }m\neq2a-1,\\
                K_{\frac{2a-m-1}{2}}(|p_{b}|) & \sim-\ln\left(\frac{|p_{b}|}{2}\right)-\zeta, & \text{when }m=2a-1,\label{eq:Asymp_Mod_Bess_Second_Kind_0}
        \end{align}
        where $\zeta$ is the Euler-Mascheroni constant. Thus, the wave function
        diverges at $p_{b}\to0$ only if $m=2a-1$. For values of
        $m\neq2a-1$, the wave function reduces to 
        \begin{equation}
                \bar{\psi}_{m}\propto p_{c}^{-1+a+\frac{m}{4}}.\label{eq:wave-func-p_c-usuall}
        \end{equation}
        Based on all the above observations, we find that the wave function diverges in the limit $p_{c}\to0$ where classical singularity resides. This means that the classical singularity
        is not resolved in the context of the standard quantization that we performed above.
\subsection{Kretschmann scalar}     
    This observations of the previous subsection can be made more clear by computing Riemann invariants.
        These quantities, particularly the Kretschmann scalar $K=R_{\mu\nu\alpha\beta}R^{\mu\nu\alpha\beta}$, are coordinate invariant measures of whether a singularity exists in some region, given that the said region is not an infinite affine parameter away. In order to see the fate of the singularity in the quantum regime we compute the expectation value of $K$ with respect to the quantum state \eqref{Wave.Fun.Sol}. The classical $K$ in terms of the Ashtekar-Barbero variables for our model is expressed as \cite{Rastgoo:2022mks}
        \begin{equation}
                K=\frac{12}{\gamma^4}\frac{(b^2+\gamma^2)^2}{p_c^2}.
        \end{equation}
        Classically, at the event horizon, $p_c=4G^2M^2$ and $b=0$, thus in this region the Kretschmann scalar is $K=3/4G^4M^4$. On the other hand, at the center of the black hole where $p_c=0$ and $b\rightarrow\infty$, the Kretschmann scalar diverges and this signals the presence of a physical singularity in this region.
        
        After quantization, $K$ will be an operator due to its dependence on the operators $\hat{p}_c$ and $\hat{b}$.
        We can then compute its expectation value in the connection representation where $\hat{b}$ acts multiplicatively and $\hat{p}_c$ is a differential operator. Therefore, we will use the wave function   \eqref{Wave.Fun.Sol} to calculate the expectation value of the Kretschmann scalar as
        \begin{equation}
                \langle \hat{K}\rangle=\frac{\iint \psi^*(b,c) \hat{K} \psi(b,c)\ db\ dc}{\iint |\psi(b,c)|^2\ db\ dc}.\label{eq:Kretsch_Exp_Gen}
        \end{equation}
To do this, we first calculate the denominator of \eqref{eq:Kretsch_Exp_Gen} using equation \eqref{Wave.Fun.Sol}, as
\begin{equation}
    \iint |\psi(b,c)|^2\ db\ dc=\underbrace{\int_{-\infty}^{\infty}(b^2+\gamma^2)^{m-2a}\ db}_{J_1}\underbrace{\int_{-\infty}^{\infty}c^{-m/2-2a}\ dc}_{J_2}.\label{eq:Norm_wave_func_usuall}
\end{equation}
We compute each of the integrals labeled $J_1$ and $J_2$ by writing them as $\lim_{D\to\infty}\int_{-D}^{D}\cdots$ for some parameter $D$.
Then, the first integral, $J_1$, yields
\begin{equation}
    J_1=\frac{\sqrt{\pi } \left(\gamma ^2\right)^{-2 a+m+1/2} \Gamma \left(2 a-m-\frac{1}{2}\right)}{\Gamma (2 a-m)},\label{eq:J_1_integral}
\end{equation}
with the integration conditions $4a-2m>1$ and $\gamma^2>0$. 
The second integral, $J_1$, before taking the limit $D\to\infty$ becomes
\begin{equation}
    J_2=\int_{-D}^{D}c^{-m/2-2a}\ dc=\frac{2 (-D)^{-2 a-\frac{m}{2}+1}\left(1+e^{\frac{1}{2} i \pi  (4 a+m)}\right)}{4 a+m-2},\label{eq:J_2_integral}
\end{equation}
with the integration conditions $2a+m/2<1$ and $D\geq0$. Additionally, we require that the integral of the squared norm of the wave function is real and positive. Therefore, we define a new quantum number $n=2a+m/2$, in such a way that, based on the Euler formula, $e^{in\pi}=(-1)^n$ with $n$ being a negative even integer $(n=0,-2,-4,\cdots)$. In this way \eqref{eq:J_2_integral} reduces to
\begin{equation}
    J_2=\frac{2\left(-D\right)^{1-n}}{n-1}.\label{eq:J_2_integral_solution}
\end{equation}
It is now clear that $ \lim_{D\to\infty} J_2$ does not converge.
One also might be concerned about the limits of integration going from $-\infty$ to $+\infty$ given that the range of values of $b$ and $c$. This certainly needs a more careful analysis of the mathematical structure of the state space of the model but for now we can remedy it by defining functions of $b$ and $c$ in a piecewise manner such that they are defined to be identically zero outside their given range. But what we can expect is that this issue with the Hilbert space in the standard quantization, not surprisingly, will persist, as in the well-known case of the Wheeler-Dewitt theory where the kinematical Hilbert space is ill-defined. Nevertheless, if we take the above result at face value, it shows that $\langle K \rangle$ does not remain finite in the entire interior region. Note that these issues of the Hilbert space will dispappear in the minimal uncertainty case that we consider in the next section.

Combining the integration conditions of \eqref{eq:Wav.Sol.Mod.Bess.Two.Without.GUP}, \eqref{eq:J_1_integral} and \eqref{eq:J_2_integral}, we find a restriction for the values that $m$ as $m <1/3$. Also, the relation $m=2(n-2a)$, implies that the quantum number $m$ must be a negative integer, $m<0$, which is already satisfied in our conditions. If, for example, we choose the factor ordering parameter $a = 1/2$ corresponding to a symmetric Hamiltonian constraint operator, we would have have $m=2(n-1)$.

Continuing with the computations of $\langle K \rangle$, the numerator of \eqref{eq:Kretsch_Exp_Gen} using the conditions above becomes
        \begin{align}
                \iint \psi^*(b,c) \hat{K} \psi(b,c)\ db\ dc=\frac{12}{\gamma^4}\lim_{D\to\infty}\int_{-D}^{D}(b^2+\gamma^2)^{m-2a+2}\ db\lim_{D\to\infty}\int_{-D}^{D}c^{-m/4-a}\frac{1}{\hat{p}_c^2}c^{-m/4-a}\ dc.\label{eq.Expect_Value_K_Withou_Norm}
        \end{align}
To calculate the integral in $c$, we need to know how the operator $1/\hat{p}_c^2$ acts on $c^{-m/4-a}$. For this purpose, we introduce an arbitrary function $\varphi(c)$ such that $\hat{p}_c^2 \varphi(c) = c^{-m/4-a}$.
        Solving for $\varphi(c)$, we get
        \begin{equation}
                \varphi(c) 
                = -\frac{1}{4G^2\gamma^2}\int_{0}^{c}dx\int_{0}^{x}y^{-m/4-a}\ dy
                = -\frac{4c^{2-a-m/4}}{G^2\gamma^2(4a+m-4)(4a+m-8)},\label{eq:New_function_varphi}
        \end{equation}
with the condition $m<4-4a$. Therefore, with this result, we obtain for the integral in $c$ in \eqref{eq.Expect_Value_K_Withou_Norm},
\begin{equation}
    \int_{-D}^{D}c^{-m/4-a}\frac{1}{\hat{p}_c^2}c^{-m/4-a}\ dc=-\frac{8\left(1+e^{\frac{1}{2} i \pi  (4 a+m)}\right) (-D)^{-2 a-\frac{m}{2}+3}}{G^2\gamma^2(4a+m-4)(4a+m-6)(4a+m-8)},\label{eq:c_Integral_usuall}
\end{equation}
where the integration conditions are $m < 6 - 4a$ and $D\geq 0$. According to the criteria we considered in obtaining \eqref{eq:J_2_integral_solution}, the result in \eqref{eq:c_Integral_usuall} reduces to
\begin{equation}
    \int_{-D}^{D}c^{-m/4-a}\frac{1}{\hat{p}_c^2}c^{-m/4-a}\ dc=-\frac{2(-D)^{3-n}}{G^2\gamma^2(n-2)(n-3)(n-4)}.\label{eq:c_solution_usuall}
\end{equation}
But as we saw before, $n$ must be an even negative integer. This means that  \eqref{eq:c_solution_usuall} is negative for any value of $n$ that meets the aforementioned conditions.
On the other hand, the integral of $b$ in \eqref{eq.Expect_Value_K_Withou_Norm} is computed to be
\begin{equation}
    \lim_{D\to\infty}\int_{-D}^{D}(b^2+\gamma^2)^{m-2a+2}\ db=\frac{\sqrt{\pi}\gamma^{5-4a+2m}\Gamma\left(2a-m-\frac{5}{2}\right)}{\Gamma(2a-m-2)},\label{eq:b_Integral_usuall}
\end{equation}
and converges under the condition $m<2a-5/2$.

From the integration conditions, the possible values for the factor ordering parameter $a$ are restricted. In particular, we are interested in $a=1/2$, which symmetrizes the Hamiltonian. For this case, from \eqref{eq:New_function_varphi}, we have $m<2$, from \eqref{eq:c_Integral_usuall}, $m<4$, and from \eqref{eq:b_Integral_usuall}, $m<-3/2$. These together imply that the values of the quantum number $m$ are restricted to $m<-3/2$. On the other hand, this restriction to $m$, is in accordance with the condition $m=2(n-1)$, where $n$  is a negative even integer.

Combining all the results, we obtain the expectation value of the Kretschmann scalar from \eqref{eq:Kretsch_Exp_Gen}. For $a=1/2$, it takes the form
\begin{equation}
    \langle \hat{K}\rangle_n=-\frac{48 D^2 (1-2 n) (2-2 n) (n-1)}{\gamma ^2 G^2 (n-4) (n-3) (n-2) (-2 (2 n-2)-3) (-2 (2 n-2)-1)}.\label{eq:Expec_value_K_n_witou_GUP}
\end{equation}
From this expression, it is clear that as $D\to\infty$, the expectation value of $K$ diverges. It is only finite, and negative (see Table \ref{tab:Values_of_K_several_n}), when the constant $D$ is finite.
\begin{table}[htbp]
\captionsetup{format= hang, justification=raggedright, font=footnotesize}
        \centering
        \begin{tabular}{cc}
                \toprule
                $n$ &                            $\langle \hat{K}\rangle_n$                                             \\
                \midrule
                0   &                             $-\frac{4 D^2}{3 \gamma ^2 G^2}$                                       \\
                -2   &                             $-\frac{4 D^2}{11 \gamma ^2 G^2}$                       \\
                -4   &                             $-\frac{450 D^2}{2261 \gamma ^2 G^2}$                        \\
                -6  &                             $-\frac{1274 D^2}{10125 \gamma ^2 G^2}$                        \\
                \bottomrule
        \end{tabular}
        \caption{Values of $\langle \hat{K}\rangle_n$, in \eqref{eq:Expec_value_K_n_witou_GUP}, for different negative even integer $n$ and for $a=1/2$.}
        \label{tab:Values_of_K_several_n}
\end{table}

\section{Quantization including a minimal length\label{Sec:Quantize-GUP}}

\subsection{States}     

We now consider a deformation of the standard uncertainty relations to accommodate a minimal uncertainty in the variables $p_b$ and $p_c$ according to
\begin{align}
        \Delta b\Delta p_{b}\geq & \frac{G\gamma}{2}\left[1+\beta_{b}(\Delta b)^{2}\right],\label{eqn:unc_b_pb}\\
        \Delta c\Delta p_{c}\geq & G\gamma\left[1+\beta_{c}(\Delta c)^{2}\right],\label{eqn:unc_c_pc}
\end{align}
which correspond to minimal uncertainties in $p_{b}$ and $p_{c}$
of the order $G\gamma\sqrt{\beta_{b}}$ and $2G\gamma\sqrt{\beta_{c}}$,
respectively. Therefore, $\beta_{b}$ and $\beta_{c}$ effectively
define the magnitude of the minimal uncertainty effects. 
To implement such a modification, we consider the following deformed algebra
\begin{align}
        \left[b,p_{b}\right]= & iG\gamma\left(1+\beta_{b}b^{2}\right),\label{eq:Modified_Commut_Relation_b_pb}\\
        \left[c,p_{c}\right]= & i2G\gamma\left(1+\beta_{c}c^{2}\right).\label{eq:Modified_Commut_Relation_c_pc}
\end{align}
Clearly, for $\beta_{b}=0=\beta_{c}$ we recover the standard
commutation relations \eqref{eq:Usual_Commut_Relation}.


Following standard procedures, it is convenient to introduce a new set of variables, $b_0$ and $c_0$, conjugate to $p_b$ and $p_c$, respectively, that is satisfying the following relations \cite{Pedram:2011aa,Bosso:2023sxr}
\begin{align}
        [b_{0},p_{b}]= & iG\gamma, & [c_{0},p_{c}]= & 2iG\gamma.
\end{align}
Working in the representation in which $b_{0}$ and $c_{0}$ are multiplicative
operators, the momentum operators are represented as $p_{b}=-iG\gamma\partial/\partial b_{0}$
and $p_{c}=-2iG\gamma\partial/\partial c_{0}$.
Furthermore, from \eqref{eq:Modified_Commut_Relation_b_pb} and \eqref{eq:Modified_Commut_Relation_c_pc}, we can find the following relations
\begin{align}
        b & =\frac{1}{\sqrt{\beta_{b}}}\tan\left(\sqrt{\beta_{b}}b_{0}\right), & c & =\frac{1}{\sqrt{\beta_{c}}}\tan\left(\sqrt{\beta_{c}}c_{0}\right),\label{eq:Self-adjoint_representation}
\end{align}
where the domains of $b_{0}$ and $c_{0}$ are restricted to $-\pi\sqrt{\beta_{b}}/2<b_{0}<\pi\sqrt{\beta_{b}}/2$ and $-\pi\sqrt{\beta_{c}}/2<c_{0}<\pi\sqrt{\beta_{c}}/2$ respectively
\cite{Bosso:2023sxr}.


It is now possible to rewrite the Hamiltonian \eqref{eq:Quantum-Hamiltonian}
in terms of the new canonically conjugate variables $(b_{0},c_{0})$
and $(p_{b},p_{c})$.
In this way, the differential equation resulting from the action of the representation of the Hamiltonian constraint on a state in the $(b_{0},c_{0})$ space
reads 
\begin{equation}
        \left[\left(b^{2}+\gamma^{2}\right)^{1-a}\frac{\partial}{\partial b_{0}}\left(b^{2}+\gamma^{2}\right)^{a}+4bc^{1-a}\frac{\partial}{\partial c_{0}}c^{a}\right]\psi_{\text{GUP}}(b_{0},c_{0})=0.\label{Eq.Diff.Eq.Conn.Repre.GUP}
\end{equation}
Here, as in \eqref{Diff.Eq.Conn.Repr}, $a$ parametrizes a generic factor ordering, and $b$ and $c$ are given in terms of the variables $(b_{0},c_{0})$ as in \eqref{eq:Self-adjoint_representation}.
Proceeding as in the previous section, we would like to solve the
differential equation \eqref{Eq.Diff.Eq.Conn.Repre.GUP} by separation
of variables of the form $\psi_{\text{GUP}}(b_{0},c_{0})=B(b_{0})R(c_{0})$,
from which one obtains two differential equations, one for $b_{0}$
\begin{equation}
        \left(\frac{\tan^{2}\left(\sqrt{\beta_{b}}b_{0}\right)}{\beta_{b}}+\gamma^{2}\right)\frac{dB(b_{0})}{db_{0}}+\frac{\tan\left(\sqrt{\beta_{b}}b_{0}\right)}{\sqrt{\beta_{b}}}\left(2a\sec^{2}\left(\sqrt{\beta_{b}}b_{0}\right)-m\right)B(b_{0})=0,\label{eq:Solution_for_b_0_with_GUP}
\end{equation}
and one for $c_{0}$ 
\begin{align}
        4\frac{\tan\left(\sqrt{\beta_{c}}c_{0}\right)}{\sqrt{\beta_{c}}}\frac{dR(c_{0})}{dc_{0}}+\left(m+4a\sec^{2}\left(\sqrt{\beta_{c}}c_{0}\right)\right)R(c_{0})=0.\label{eq:Solution_for_C_0_with_GUP}
\end{align}
These equations can be solved, in the same manner as before, obtaining the solutions
\begin{align}
        B_{m}(b_{0})= & C_{1_m}\left[\sec^2(\sqrt{\beta_b}b_0)\right]^{-\frac{m}{2(1-\beta_b\gamma^2)}}\left(\frac{\tan^2{(\sqrt{\beta_b}b_0)}}{\beta_b}+\gamma^2\right)^{-a+\frac{m}{2(1-\beta_b\gamma^2)}},\label{Eq:B(b_0)_GUP_reduc}\\
        R_{m}(c_{0})= & C_{2_m}\left[\sec^2(\sqrt{\beta_c}c_0)\right]^{\frac{m}{8}}\left(\frac{\tan{(\sqrt{\beta_c}c_0)}}{\sqrt{\beta_c}}\right)^{-a-\frac{m}{4}}.\label{Eq.R(c_0)_GUP_reduc}
\end{align}
Furthermore, it is worth noticing that, again in the limits $\beta_b \to 0$ and $\beta_c \to 0$, we recover the ordinary solutions in \eqref{eq:Usual_b_solution} and \eqref{eq:Usual_c_solution}.

As before, the momentum space wave functions are derived from the Fourier transform 
\begin{equation}
        \bar{\psi}_{m}^{GUP}(p_{b},p_{c})=\int_{-\frac{\pi}{2\sqrt{\beta_{b}}}}^{\frac{\pi}{2\sqrt{\beta_{b}}}}\int_{-\frac{\pi}{2\sqrt{\beta_{c}}}}^{\frac{\pi}{2\sqrt{\beta_{c}}}}e^{-ip_{b}b_{0}}e^{-ip_{c}c_{0}}\ \psi_{m}^{GUP}(b_{0},c_{0})\ db_{0}\ dc_{0},\label{Four.Trans.GUP}
\end{equation}
or equivalently 
\begin{align}
        \bar{\psi}_{m}^{GUP}\left(p_{b},p_{c}\right)= & \int_{-\frac{\pi}{2\sqrt{\beta_{b}}}}^{\frac{\pi}{2\sqrt{\beta_{b}}}}e^{-ip_{b}b_{0}}\ B_{m}(b_{0})\ db_{0}\int_{-\frac{\pi}{2\sqrt{\beta_{c}}}}^{\frac{\pi}{2\sqrt{\beta_{c}}}}e^{-ip_{c}c_{0}}\ R_{m}(c_{0})\ dc_{0},\nonumber \\
        = & \bar{B}_{m}(p_{b})\bar{R}_{m}(p_{c}),\label{Four.Trans.GUP.saparated}
\end{align}
where $\bar{B}(p_{b})$ represents the ordinary Fourier transform of
$B_{m}(b_{0})$, and similarly for $\bar{R}(p_{c})$, i.e.,
\begin{align}
        \bar{B}_{m}(p_{b})= & \int_{-\frac{\pi}{2\sqrt{\beta_{b}}}}^{\frac{\pi}{2\sqrt{\beta_{b}}}}e^{-ip_{b}b_{0}}\ B_{m}(b_{0})\ db_{0},\label{Eq:B(p_b)_GUP}\\
        \bar{R}_{m}(p_{c})= & \int_{-\frac{\pi}{2\sqrt{\beta_{c}}}}^{\frac{\pi}{2\sqrt{\beta_{c}}}}e^{-ip_{c}c_{0}}\ R_{m}(c_{0})\ dc_{0}.\label{Eq.R(p_c)_GUP}
\end{align}
For an arbitrary factor ordering parameter $a$, \eqref{Eq.R(p_c)_GUP} becomes
\begin{multline}
    \bar{R}_{m}(p_{c})= i (-1)^{-\frac{m}{8}} 2^{\frac{m}{4}-1}\beta_c^{\frac{1}{8} (4 a+m-4)} e^{-\frac{1}{8} i \pi  \left(4 a+m+\frac{4 p_c}{\sqrt{\beta_c}}\right)}\left\{\Gamma \left(-a-\frac{m}{4}+1\right) e^{\frac{1}{8} i \pi  \left(m+\frac{4 p_c}{\sqrt{\beta_c}}\right)}\right.\\
    \left.\times\left[\Gamma \left(\frac{1}{8} \left(m-\frac{4 p_c}{\sqrt{\beta_c}}\right)\right) \, _2\tilde{F}_1\left(-a,\frac{1}{8} \left(m-\frac{4 p_c}{\sqrt{\beta_c}}\right);-a-\frac{m}{8}-\frac{p_c}{2 \sqrt{\beta_c}}+1;-1\right)\right.\right.\\
    \left.\left.+\Gamma \left(\frac{1}{8} \left(m+\frac{4 p_c}{\sqrt{\beta_c}}\right)\right) \, _2\tilde{F}_1\left(-a,\frac{1}{8} \left(m+\frac{4 \text{pc}}{\sqrt{\text{$\beta $c}}}\right);-a-\frac{m}{8}+\frac{\text{pc}}{2 \sqrt{\text{$\beta $c}}}+1;-1\right)\right]\right.\\
    \left.-\Gamma (a+1)\left[e^{\frac{i \pi  m}{4}} \Gamma \left(\frac{1}{8} \left(m-\frac{4 p_c}{\sqrt{\beta_c}}\right)\right) \, _2\tilde{F}_1\left(a+\frac{m}{4},\frac{1}{8} \left(m-\frac{4p_c}{\sqrt{\beta_c}}\right);a+\frac{1}{8} \left(m-\frac{4 p_c}{\sqrt{\beta_c}}+8\right);-1\right)\right.\right.\\
    \left.\left.+e^{\frac{1}{4} i \pi  \left(m+\frac{4 \text{pc}}{\sqrt{\text{$\beta $c}}}\right)} \Gamma \left(\frac{1}{8} \left(m+\frac{4 \text{pc}}{\sqrt{\text{$\beta $c}}}\right)\right) \, _2\tilde{F}_1\left(a+\frac{m}{4},\frac{1}{8} \left(m+\frac{4 \text{pc}}{\sqrt{\text{$\beta $c}}}\right);a+\frac{1}{8} \left(m+\frac{4 \text{pc}}{\sqrt{\text{$\beta $c}}}+8\right);-1\right)\right]\right\},\label{eq:R(pc)_arbitrary_a}
\end{multline}
with the conditions $\beta_c>0$ and $a+\frac{m}{4}<1$. Here, $_2\tilde{F}_1(A,B;C;z)=\left._2 F_1(A,B;C;z)\right/ \Gamma(C)$ is the regularized hypergeometric function. The hypergeometric function $_2 F_1(A,B;C;z)$, converges at $z = \pm 1$ if and only if $C>A+B$. This condition applied to \eqref{eq:R(pc)_arbitrary_a} implies $m<4$. We can define additional convergence conditions based on the definitions of the Gamma function $\Gamma(z)$, where $z$ takes on non-zero values, excluding negative integers. Thus with this condition, from $\Gamma \left(-a-\frac{m}{4}+1\right)$, we obtain $-a-\frac{m}{4}+1\neq -n$, with $n=0,1,2,\cdots$. Therefore $m\neq4(n-a+1)$. Particularly, for $a=1/2$, which symmetrizes the Hamiltonian in \eqref{Eq.Diff.Eq.Conn.Repre.GUP}, the condition for $m$ becomes in $m<4(n+1/2)$, which, for $n=0$, satisfy the integration condition in \eqref{eq:R(pc)_arbitrary_a}. Similarly, from $\Gamma \left(\frac{1}{8} \left(m-\frac{4 p_c}{\sqrt{\beta_c}}\right)\right)$ and $\Gamma \left(\frac{1}{8} \left(m+\frac{4 p_c}{\sqrt{\beta_c}}\right)\right)$ we find 
\begin{equation}
    p_{c}\neq\pm\frac{\sqrt{\beta_{c}}}{4}(m+8n)\label{pc-val-min-n}
\end{equation}
respectively for plus and minus signs. Additionally, from $\Gamma (a+1)$, we deduce $a\neq -n-1$, which implies $a>-n-1$, consistent with the condition set for $a=1/2$. 

For the specific case of a symmetric Hamiltonian constraint, $a=1/2$, the expression \eqref{eq:R(pc)_arbitrary_a} reduces to
\begin{multline}
    \bar{R}_{m}(p_{c})
        = \frac{2^{\frac{m}{4}-2}C_{2_m} \Gamma \left(\frac{1}{8} \left(m-\frac{4 p_c}{\sqrt{\beta_c}}\right)\right)}{\beta_c}\left[2 \left(i \sqrt{\beta_c}\right)^{\frac{m}{4}+\frac{3}{2}} \left(\left(-1\right)^{\frac{m+2}{4}}-1\right) \Gamma \left(\frac{1}{2}-\frac{m}{4}\right)\right.\\
        \left.\times\, _2\tilde{F}_1\left(-\frac{1}{2},\frac{1}{8} \left(m-\frac{4 p_c}{\sqrt{\beta_c}}\right);\frac{1}{8} \left(-m-\frac{4p_c}{\sqrt{\beta_c}}+4\right);-1\right)\right.\\
        \left.- \sqrt{\pi } \beta_c^{\frac{m+4}{8}} e^{-\frac{i \pi  p_c}{2 \sqrt{\beta_c}}} \left(\sqrt{i \sqrt{\beta_c}}+\sqrt{-i \sqrt{\beta_c}} e^{\frac{i \pi p_c}{\sqrt{\beta_c}}}\right)\right.\\
        \left.\times\, _2\tilde{F}_1\left(\frac{m+2}{4},\frac{1}{8} \left(m-\frac{4 p_c}{\sqrt{\beta_c}}\right);\frac{1}{8} \left(m-\frac{4 p_c}{\sqrt{\beta_c}}+12\right);-1\right)\right].\label{eq:Rm(pc)_GUP_a_un_medio}
\end{multline}
In this case, the quantum number $m$ is restricted to $m<2$. 

Moreover, when $p_{c}\rightarrow0$, \eqref{eq:R(pc)_arbitrary_a} reduces to
\begin{equation}
    \bar{R}_{m}(0) = \frac{\pi\beta_c^{\frac{1}{8} (4 a+m-4)}  \sec \left(\frac{\pi  a}{2}\right) \sin \left(\frac{\pi  m}{8}\right) \Gamma \left(\frac{m}{8}\right) \sec \left(\frac{1}{8} \pi  (4 a+m)\right)}{\Gamma \left(\frac{1}{2}-\frac{a}{2}\right) \Gamma \left(\frac{1}{8} (4 a+m+4)\right)}.\label{eq:wave-function-GUP-p_c=00003D0}
\end{equation}
One can see that this result differs from the one obtained in the standard quantization, \eqref{eq:wave-func-p_c-usuall}, in that the above is finite under certain conditions (see below) while \eqref{eq:wave-func-p_c-usuall} diverges at $p_c = 0$.
This is a consequence of the modification of the algebra \eqref{eq:Modified_Commut_Relation_c_pc}, effectively introducing a minimal uncertainty in $p_{c}$.
Furthermore, it is easy to see that when $\beta_{c} \to 0$, the wave function in \eqref{eq:wave-function-GUP-p_c=00003D0} diverges for the condition $a+m/4<1$, as expected. From conditions on the convergence of \eqref{eq:wave-function-GUP-p_c=00003D0}, we can derive constraints on the possible values of the quantum number $m$. Convergence of $\Gamma \left(\frac{m}{8}\right)$, yields $m\neq-8n$ with an integer $n$. The secant function is finite for $m\neq-4(2n+1+a)$. Under these conditions, we observe that $1/\Gamma \left(\frac{1}{8} (4 a+m+4)\right)$ cannot be zero as desired. For $a = 1/2$, these conditions mean that $m$ cannot take values $0, -6, -8, -14, -16, -22, \cdots$, for $n = 0, 1, 2, \cdots$, respectively. This indicates that the wave function at $p_c = 0$ is finite and nonzero, with the specified constraints above.


In a similar manner, we can obtain $\bar{B}_{m}$ from \eqref{Eq:B(p_b)_GUP} for an arbitrary $a$

\begin{equation}
    \bar{B}_{m}(p_{b})=\int_{-\frac{\pi }{2 \sqrt{\beta_b}}}^{\frac{\pi }{2 \sqrt{\beta_b}}} e^{-i b_0 p_b} \left[\sec^2(\sqrt{\beta_b}b_0)\right]^{-\frac{m}{2(1-\beta_b\gamma^2)}}\left(\frac{\tan^2{(\sqrt{\beta_b}b_0)}}{\beta_b}+\gamma^2\right)^{-a+\frac{m}{2(1-\beta_b\gamma^2)}} \, db_0.\label{eq:Integral_B(b0)_arbit_a}
\end{equation}
Such integral for $a=1/2$ is
\begin{align}
    \bar{B}_{m}(p_{b})=&\frac{C_{1_m}\pi \beta_b^{\frac{m}{2 \beta_b \gamma ^2-2}}}{2}  \, _3\tilde{F}_2\left(1,\frac{3}{2},\frac{m}{2 \beta_b \gamma ^2-2}+\frac{1}{2};\frac{3}{2}-\frac{p_b}{2 \sqrt{\beta_b}},\frac{1}{2} \left(\frac{p_b}{\sqrt{\beta_b}}+3\right);1-\beta_b \gamma ^2\right),\label{eq:B(pb)_for_a_1/2_GUP}
\end{align}
with the condition $\beta_b \gamma ^2\geq 0$. The hypergeometric function in \eqref{eq:B(pb)_for_a_1/2_GUP} can converge in two cases: either it is an infinite series in which case it converges under the condition 
\begin{equation}
    0 < \beta_b \gamma^2 < 2,
    \label{eqn:condition1}
\end{equation} 
or it is finite, which automatically guarantees its convergence, in which case the finiteness condition is that either of the three first arguments is a non-positive integer. This leads to the condition
\begin{equation}
    - \frac{m}{2(1 - \beta_{b}\gamma^{2})} + \frac{1}{2} = - n \qquad \Rightarrow \qquad
    m = 2 \left(n + \frac{1}{2}\right) (1 - \beta_{b}\gamma^{2}).
    \label{eqn:condition2}
\end{equation}
In the latter case, either $n$ can be considered an integer quantum number, or equivalently $m$ can be considered a real valued quantum number. Notice that the two above conditions \eqref{eqn:condition1} and \eqref{eqn:condition2} are independent. It is important to mention that $m < 0$, according to the convergence conditions discussed so far. Therefore, we infer from \eqref{eqn:condition2} that
\begin{equation}
    1 - \beta_{b}\gamma^{2}<0.\label{enq:Condition_beta_b_gamma}
\end{equation} 
On the other hand, \eqref{eq:B(pb)_for_a_1/2_GUP} gives us conditions for $p_b=0$. Notice that according the line just after \eqref{eq:Black.Hole.Metric.In.Ashtekar.Varbl}, classically singularity, is located where both $p_c,\,p_b\to0$. So, it is also important to consider the case where $p_b\to 0$. For $p_b=0$ from \eqref{eq:B(pb)_for_a_1/2_GUP} we obtain
\begin{equation}
    \bar{B}_{m}(0)=2\beta_b^{\frac{m}{2 \beta_b \gamma ^2-2}} \, _2F_1\left(1,\frac{m}{2 \beta_b \gamma ^2-2}+\frac{1}{2};\frac{3}{2};1-\beta_b \gamma ^2\right).
\end{equation}
As we mentioned earlier, the absolute convergence of the hypergeometric function $\, _2F_1(A,B;C;z)$ occurs when $C > A + B$ in the unit circle $|z|=1$. In our case, this condition is satisfied $|1-\beta_b\gamma^2|=1$, or equivalently, $1-\beta_b\gamma^2=-1$, which is in line with \eqref{enq:Condition_beta_b_gamma}. Thus we have arrived at the conclusion 
\begin{equation}
    \beta_b=\frac{2}{\gamma^2}.\label{eq:Relation_beta_b_gamma}
\end{equation}
This very interesting observation sets a strict relation between a free parameter in LQG, namely the Barbero-Immirzi parameter $\gamma$, and one of the minimum uncertainty free parameters $\beta_b$. Notice that this is the direct consequence of demanding the convergence of the wave function, hence asking the states to be square-integrable, everywhere, particularly where used to be the classical singularity. 

\begin{figure}[t!]
\captionsetup{format= hang, justification=raggedright, font=footnotesize}
        \begin{centering}
        \includegraphics[scale=0.45]{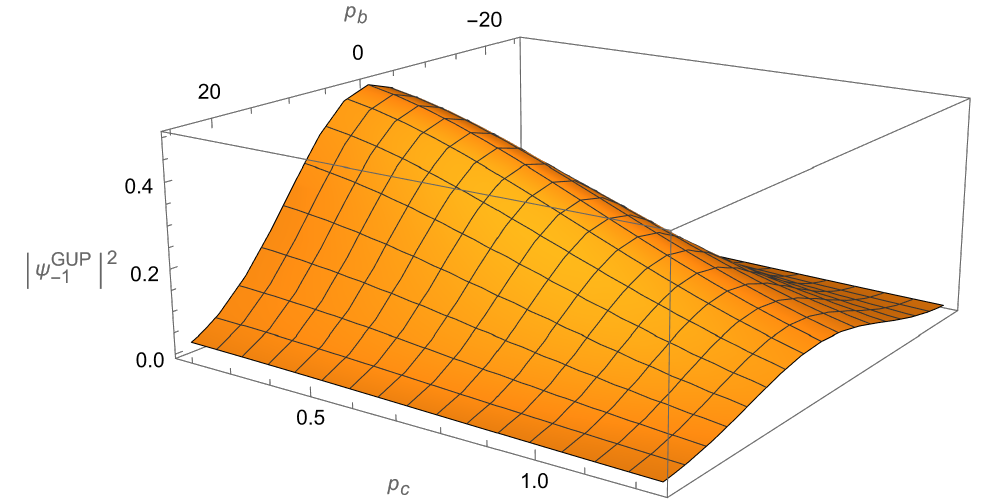}~~~~\includegraphics[scale=0.3]{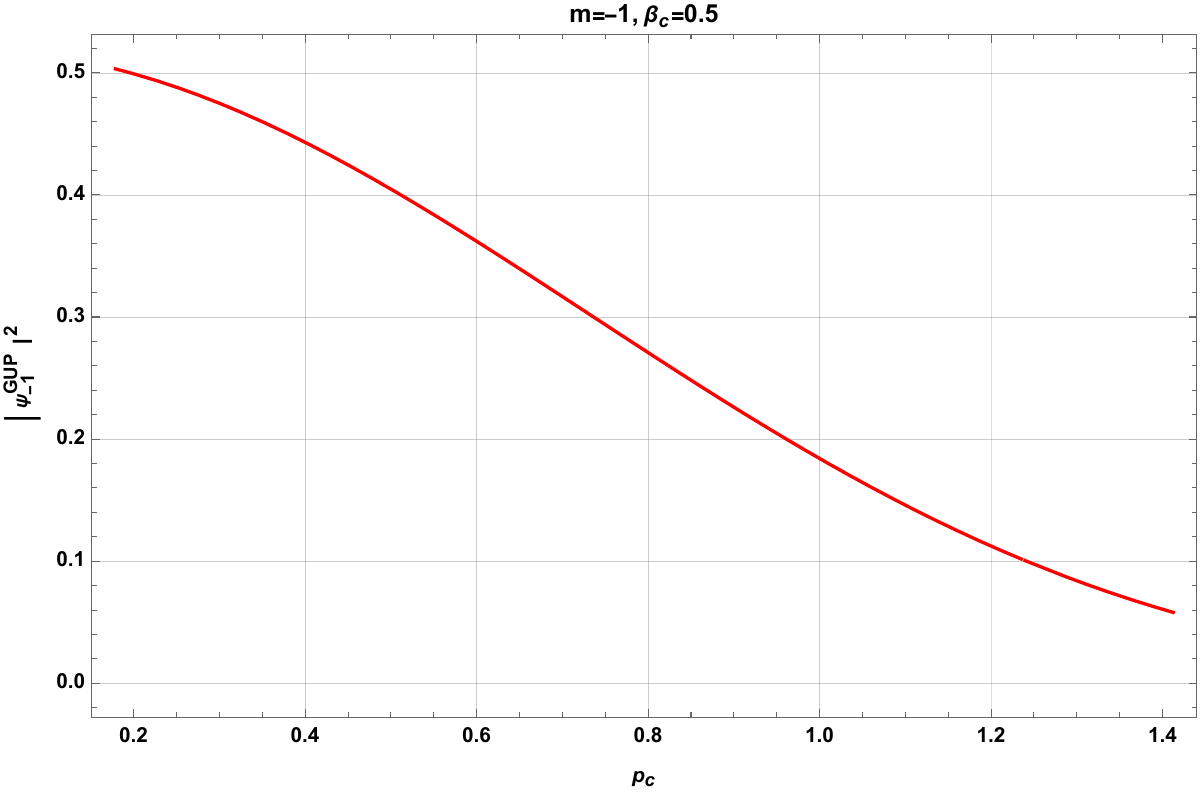}
        \par\end{centering}
        \caption{Modulus squared of the wave function $|\psi_{-1}|^{2}$.}
        \label{fig:Plot_3D_Gup_4}
\end{figure}

\begin{figure}[t!]
\captionsetup{format= hang, justification=raggedright, font=footnotesize}
        \begin{centering}
        \includegraphics[scale=0.4]{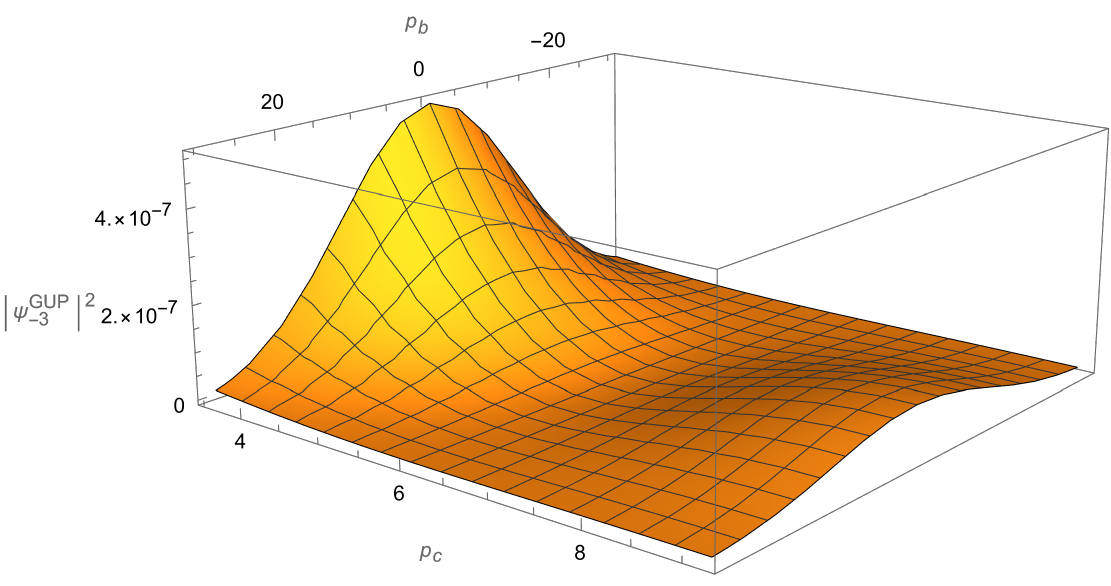}~~~~\includegraphics[scale=0.3]{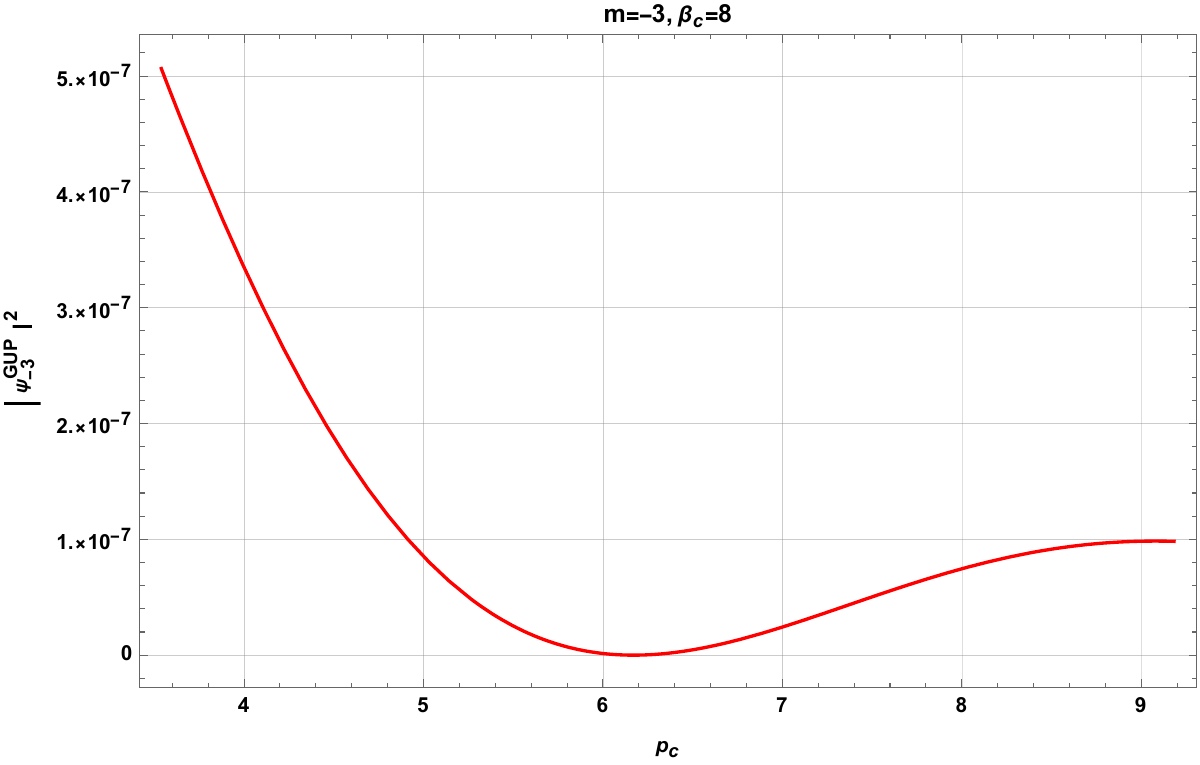}
        \par\end{centering}
        \caption{Modulus squared of the wave function $|\psi_{-3}|^{2}$.}
        \label{fig:Plot_3D_Gup_12}
\end{figure}

\begin{figure}[t!]
\captionsetup{format= hang, justification=raggedright, font=footnotesize}
        \begin{centering}
        \includegraphics[scale=0.4]{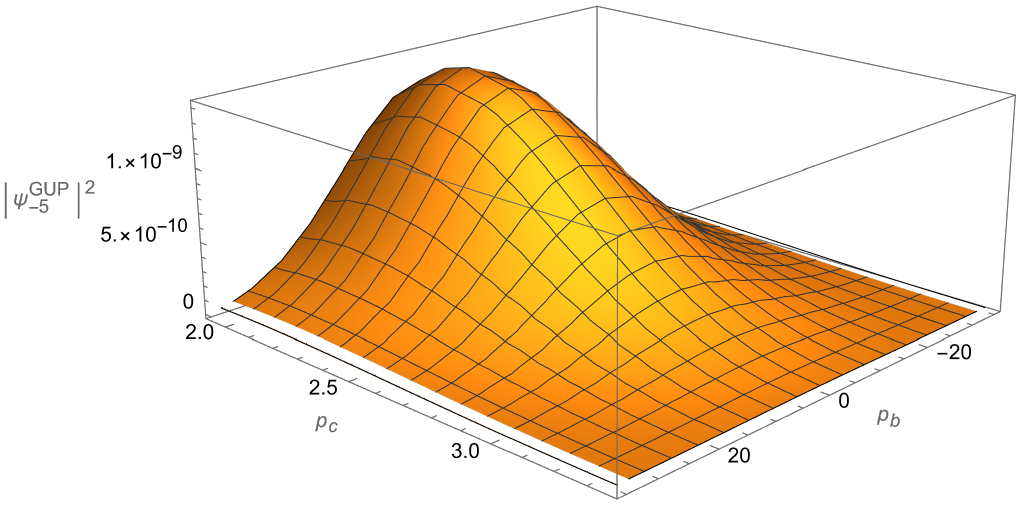}~~~~\includegraphics[scale=0.3]{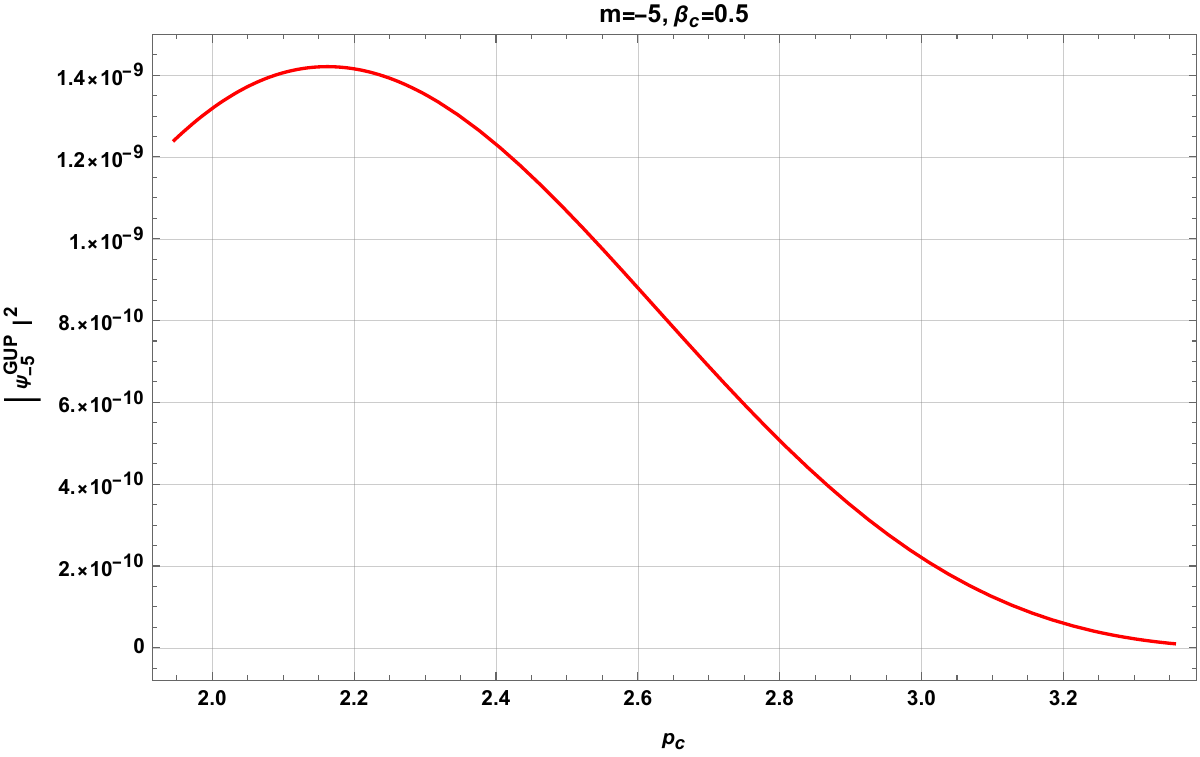}
        \par\end{centering}
        \caption{Modulus squared of the wave function $|\psi_{-5}|^{2}$.}
        \label{fig:Plot_3D_Gup_20}
\end{figure}

Replacing $\beta_b$ from \eqref{eq:Relation_beta_b_gamma} into \eqref{eqn:condition2}, we obtain
\begin{equation}
    m = -2(n+\frac{1}{2}),\label{eq:Relation_m_with_n}
\end{equation}
which implies that the possible values of $m$ are negative odd integers. This is another interesting restriction, now on the values that the quantum number $m$ (governed by $n$), which is dictated by the convergence of $\bar{R}_m$ in \eqref{eq:wave-function-GUP-p_c=00003D0} and therefore of \eqref{eq:Rm(pc)_GUP_a_un_medio} for the case of a symmetric Hamiltonian constraint, $a = 1/2$. The above analysis lets us conclude that the wave function remains finite in the whole interior region, and particularly at $p_c,\,p_b\to0$ where used to be the location of the classical singularity, when conditions \eqref{eq:Relation_beta_b_gamma} and \eqref{eq:Relation_m_with_n} are met. Figures \ref{fig:Plot_3D_Gup_4}-\ref{fig:Plot_3D_Gup_20} show the plots of the modulus squared of the wave functions for three different values of $m$.

Another interesting result is obtained by replacing $m$ from \eqref{eq:Relation_m_with_n} into \eqref{pc-val-min-n}, yielding 
\begin{equation}
p_{c}\neq\pm\frac{\sqrt{\beta_{c}}}{4}(6n-1)
\end{equation}
However, remembering the condition $n=0,1,2,\cdots$ we derived in the paragraph just above \eqref{pc-val-min-n}, this becomes
\begin{equation}
p_{c}\neq\pm\frac{\sqrt{\beta_{c}}}{4}.
\end{equation}
Since $p_{c}$ corresponds to the square of the radius of the 2-spheres in the interior of the black hole, assuming that initially $p_{c}>0$, and its evolution is continuous, we infer from the previous equation
\begin{equation}
p_{c}>\frac{\sqrt{\beta_{c}}}{4}.
\end{equation}
This shows that the minimal uncertainty approach enforces a minimum radius on the 2-spheres in the black hole, which is related to the quantum parameter $\beta_c$. Notice that this is a universal bound, meaning that it is independent of the mass of the black hole and purely a quantum effect, which vanishes for $\beta_c=0$. This can also be seen as a weaker argument for the resolution of the singularity in this approach. A stronger indication of the resolution of the singularity is derived in what follows, by actually showing that the expectation value of the Kretschmann scalar $\langle \hat{K}\rangle$ remains finite everywhere in the interior, even if we ignore the above condition and compute $\langle \hat{K}\rangle$ at $p_c=0$. Clearly then $\langle \hat{K}\rangle$ is also finite for $p_{c}>\frac{\sqrt{\beta_{c}}}{4}$.

\subsection{Kretschmann scalar}     
As mentioned above, we can calculate the expectation value of the Kretschmann scalar \eqref{eq:Kretsch_Exp_Gen} in a similar way as in the standard case to obtain more insight about the fate of singularity in this model. For simplicity, and because it is complicated to represent $\hat{b}$ and $\hat{c}$ as differential operators due to their classical forms \eqref{eq:Self-adjoint_representation}, we choose to calculate $\langle \hat{K}\rangle$ in the connection representation. This is why we also computed this expectation value in \eqref{eq:Expec_value_K_n_witou_GUP} in the previous quantization scheme in the same representation, so that we are able to compare the results from these two quantization approaches. Here also, we first consider the general factor ordering case with arbitrary $a$ and then specialize our results to the more interesting symmetric ordering case with $a=1/2$. Then for an arbitrary $a$, the denominator in \eqref{eq:Kretsch_Exp_Gen} in this minimal uncertainty quantization scheme becomes
\begin{equation}
    \iint|\psi_m(b_0,c_0)|^{2}\ db_0dc_0=\underbrace{\int_{-\frac{\pi}{2\sqrt{\beta_{b}}}}^{\frac{\pi}{2\sqrt{\beta_{b}}}} |B_{m}(b_{0})|^2\ db_{0}}_{I_1}\underbrace{\int_{-\frac{\pi}{2\sqrt{\beta_{c}}}}^{\frac{\pi}{2\sqrt{\beta_{c}}}} |R_{m}(c_{0})|^2\ dc_{0}}_{I_2},\label{eq:Norma_b_0_c_0_GUP}
\end{equation}
where, for an arbitrary factor ordering $a$, from \eqref{Eq:B(b_0)_GUP_reduc}-\eqref{Eq.R(c_0)_GUP_reduc}, we have
\begin{multline}
    I_1=\beta_b^{2 a+\frac{m}{\beta_b \gamma ^2-1}-\frac{1}{2}} \left(\beta_b \gamma ^2\right)^{\frac{m}{1-\beta_b \gamma ^2}-2 a}\\
    \times\left[\frac{\sqrt{\pi } \, _2F_1\left(\frac{1}{2},\frac{m}{1-\beta_b \gamma ^2}+1;-2 a+\frac{3}{2}+\frac{m}{1-\beta_b \gamma ^2};\beta_b \gamma ^2\right) \Gamma \left(2 a-\frac{1}{2}+\frac{m}{\beta_b \gamma ^2-1}\right)}{\sqrt{\frac{1}{\beta_b \gamma ^2}} \Gamma \left(2 a+\frac{m}{\beta_b \gamma ^2-1}\right)}\right.\\
    \left. +\frac{\Gamma \left(2 a+\frac{1}{2}\right) \, _2F_1\left(2 a+\frac{1}{2},2 a+\frac{m}{\beta_b \gamma ^2-1};2 a+\frac{1}{2}+\frac{m}{\beta_b \gamma ^2-1};\beta_b \gamma ^2\right) \Gamma \left(-2 a+\frac{1}{2}+\frac{m}{1-\beta_b \gamma ^2}\right)}{\left(\frac{1}{\beta_b \gamma ^2}\right)^{-\frac{m}{1-\beta_b \gamma ^2}+2 a}\Gamma \left(\frac{m}{1-\beta_b \gamma ^2}+1\right)}\right],\label{eq:I1_beta_b_GUP}
\end{multline}
with the integration conditions $\beta_b>0$, $a>-\frac{1}{4}$ and $\gamma^2>0$. From this expression, we can derive further constraints on the values of $m$ and $a$. From the convergence of Gamma functions in the numerator, i.e., $\Gamma \left(2 a-\frac{1}{2}+\frac{m}{\beta_b \gamma ^2-1}\right)$, $\Gamma \left(-2 a+\frac{1}{2}+\frac{m}{1-\beta_b \gamma ^2}\right)$, and $\Gamma \left(2 a+\frac{1}{2}\right)$, we find $m\neq(n+2a-1/2)(1-\beta_b\gamma^2)$, $m\neq-(n-2a+1/2)(1-\beta_b\gamma^2)$, and  $a\neq-(n/2+1/4)$, respectively. These restrict $a$ to be greater than $0$. For the special case of $a = 1/2$, and considering \eqref{eq:Relation_beta_b_gamma}, we find $m \neq-(n+1/2)$ and $m \neq(n-1/2)$, implying that $m$ cannot be a non-integer number, negative or positive. This same argument for $m$ can be extracted from the hypergeometric functions $\, _2F_1(A,B;C;z)$ appearing in \eqref{eq:I1_beta_b_GUP} with $C\neq-n$. On the other hand, if we consider $\frac{m}{1-\beta_b \gamma ^2}+1=-n$, the hypergeometric function
\begin{equation}
\, _2F_1\left(\frac{1}{2},\frac{m}{1-\beta_b \gamma ^2}+1;-2 a+\frac{3}{2}+\frac{m}{1-\beta_b \gamma ^2};\beta_b \gamma ^2\right),
\end{equation}
will not converge, since this condition implies $m=n+1$, contradicting \eqref{eq:Relation_m_with_n}. So we must set $\frac{m}{1-\beta_b \gamma ^2}+1\neq-n$. In addition, we can ensure the convergence of the first term in \eqref{eq:I1_beta_b_GUP} by fixing $2 a+\frac{m}{\beta_b \gamma ^2-1}=-2n$ in $\Gamma \left(2 a+\frac{m}{\beta_b \gamma ^2-1}\right)$, since $1/\Gamma(-2n)\to0$, naturally leading to \eqref{eq:Relation_m_with_n}. The other hypergeometric function in \eqref{eq:I1_beta_b_GUP},
\begin{equation}
    \, _2F_1\left(2 a+\frac{1}{2},2 a+\frac{m}{\beta_b \gamma ^2-1};2 a+\frac{1}{2}+\frac{m}{\beta_b \gamma ^2-1};\beta_b \gamma ^2\right),
\end{equation}
converges by virtue of \eqref{eq:Relation_m_with_n} already. Therefore, $I_1$ is reduced solely to
\begin{multline}
    I_1=\frac{\beta_b^{2 a+\frac{m}{\beta_b \gamma ^2-1}-\frac{1}{2}}\Gamma \left(2 a+\frac{1}{2}\right)\Gamma \left(-2 a+\frac{1}{2}+\frac{m}{1-\beta_b \gamma ^2}\right)}{\Gamma \left(\frac{m}{1-\beta_b \gamma ^2}+1\right)}\\
  \times  \, _2F_1\left(2 a+\frac{1}{2},2 a+\frac{m}{\beta_b \gamma ^2-1};2 a+\frac{1}{2}+\frac{m}{\beta_b \gamma ^2-1};\beta_b \gamma ^2\right),\label{eq:I1_beta_b_GUP_reduced}
\end{multline}
which is a finite quantity.

The other integral involving $c_0$ in \eqref{eq:Norma_b_0_c_0_GUP} is computed to be
\begin{equation}
    I_2=\frac{\Gamma \left(a+\frac{1}{2}\right)\beta_c^{\frac{1}{4} (4 a+m-2)} \Gamma \left(-a+\frac{1}{2}-\frac{m}{4}\right)}{\Gamma \left(1-\frac{m}{4}\right)},\label{eq:I2_beta_c_a_GUP}
\end{equation}
with the integration conditions are $\beta_c>0$, $4a+m<2$ and $a>-\frac{1}{2}$. The conditions extracted from $\Gamma \left(-a+\frac{1}{2}-\frac{m}{4}\right)$ and $\Gamma \left(1-\frac{m}{4}\right)$ in \eqref{eq:I2_beta_c_a_GUP} are $m\neq4(n-a+1/2)$ and $m\neq4(n+1)$, respectively, and they do not contradict \eqref{eq:Relation_m_with_n} for $a=1/2$. Also, from $\Gamma \left(a+\frac{1}{2}\right)$, it is inferred that $a\neq-n-1/2$, or equivalently, $a>0$. Thus, it is concluded that both \eqref{eq:I1_beta_b_GUP_reduced} and \eqref{eq:I2_beta_c_a_GUP} are finite (see Table \ref{tab:Values_I_1_I_2_several_m}), and thus so is \eqref{eq:Norma_b_0_c_0_GUP}, if \eqref{eq:Relation_beta_b_gamma} and \eqref{eq:Relation_m_with_n} are satisfied.

\begin{table}[htbp]
\captionsetup{format= hang, justification=raggedright, font=footnotesize}
        \centering
        \begin{tabular}{ccc}
                \toprule
                $m$ &                            $I_{1}\left(m,\gamma,a=1/2\right)$   &                      $I_2\left(m,\gamma,a=1/2\right)$                    \\
                \midrule
                -1   &                             $\frac{1.5708}{\sqrt{\beta_b}}$      &                $\frac{4.}{\sqrt[4]{\beta_c}}$                  \\
                -3   &                             $\frac{4.90874}{\beta_b^{5/2}}$     &                $\frac{1.33333}{\beta_c^{3/4}}$                  \\
                -5   &                             $\frac{16.3829}{\beta_b^{9/2}}$     &                $\frac{0.8}{\beta_c^{5/4}}$                   \\
                -7  &                             $\frac{57.0533}{\beta_b^{13/2}}$     &                $\frac{0.571429}{\beta_c^{7/4}}$                    \\
                \bottomrule
        \end{tabular}
        \caption{Values of $I_{1}$ and $I_2$, in \eqref{eq:I1_beta_b_GUP_reduced} and \eqref{eq:I2_beta_c_a_GUP} respectively, for different $m$ (or equivalently for $n$), for $a=1/2$ and with the condition $\beta_b \gamma^2=2$.}
        \label{tab:Values_I_1_I_2_several_m}
\end{table}

From \eqref{eq:I2_beta_c_a_GUP}, it can be observed, due to the integration condition $4a+m<2$, that when $\beta_c\to0$ the expression $I_2$ diverges. This is expected, as in the undeformed formalism, the wave function is not square-integrable, see \eqref{eq:Norm_wave_func_usuall}.

Now we turn to the numerator of \eqref{eq:Kretsch_Exp_Gen}, which can be written as
\begin{equation}
        \iint\psi_m^* \hat{K}\psi_m\ db_0dc_0=\frac{12}{\gamma^4}\underbrace{\int_{-\frac{\pi}{2\sqrt{\beta_b}}}^{\frac{\pi}{2\sqrt{\beta_b}}}|B_m(b_0)|^2(b^2+\gamma^2)^{2}\ db_0}_{L_1}\underbrace{\int_{-\frac{\pi}{2\sqrt{\beta_c}}}^{\frac{\pi}{2\sqrt{\beta_c}}}R^*_m(c_0)\frac{1}{\hat{p}_c^2}R_m(c_0)\ dc_0}_{L_2}.\label{eq:psiConju_K_psi}
\end{equation}
The integral $L_1$ in terms of the physical variable $b$, for an arbitrary factor ordering $a$, becomes
\begin{equation}
    L_1=\lim_{h\to\infty}\int_{-h}^{h}\left(1+\beta_bb^2\right)^{-1-\frac{m}{1-\beta_b\gamma^2}}\left(b^2+\gamma^2\right)^{-2a+2+\frac{m}{1-\beta_b\gamma^2}}\ db\label{eq:L1_integrate_in_b}
\end{equation}
For the symmetric case $a=1/2$, this can be computed to yield
\begin{equation}
    L_1=\lim_{h\to\infty}\int_{-h}^{h}\left(1+\beta_bb^2\right)^{-1-\frac{m}{1-\beta_b\gamma^2}}\left(b^2+\gamma^2\right)^{1+\frac{m}{1-\beta_b\gamma^2}}\ db=0.\label{eq:L1_integrate_in_b_a_1/2}
\end{equation}
Similar to the approach taken in the previous section, to determine $L_2$, we define an arbitrary function $\phi(c_0)$ such that $\frac{1}{\hat{p}_c^2}R_m(c_0)=\phi(c_0)$. From this, we derive
 \begin{equation}
     \phi(c_0)=-\frac{1}{4G^2\gamma^2}\int_{0}^{c_0}dx_0\int_{0}^{x_0}R_m(y_0)dy_0,
 \end{equation}
 but for simplicity, we make the change to physical variable $c=\tan{(\sqrt{\beta_c}c_0)}/\sqrt{\beta_c}$, which gives us
 \begin{equation}
     \phi(c)=-\frac{1}{4G^2\gamma^2}\int_{0}^{c}\frac{dx}{1+\beta_c x^2}\int_{0}^{x}\frac{R_m(y)}{1+\beta_c y^2}dy.\label{eq:Inte_phi_c_1}
 \end{equation}
 We evaluate the first integral on the right-hand side of \eqref{eq:Inte_phi_c_1} by expressing \eqref{Eq.R(c_0)_GUP_reduc} in terms of the variable $c$, with $a=1/2$. This yields:
 \begin{equation}
     \phi(c)=\frac{1}{(m-2)G^2\gamma^2}\int_{0}^{c}x^{\frac{1}{2}-\frac{m}{4}}\left(1+\beta_c x^2\right)^{\frac{m}{8}-2} \, _2F_1\left(\frac{1}{4},1;\frac{5}{4}-\frac{m}{8};- \beta_c x^2\right)\ dx,
 \end{equation}
 with the conditions $\beta_c>0$, $x>0$, and $m<2$ (consistent with \eqref{eq:Relation_m_with_n}), we proceed to solve this second integral by considering the power series of the hypergeometric function $\, _2F_1$. The integral is
 \begin{equation}
     \phi(c)=\frac{(-\beta_c)^{\frac{1}{8} (m-6)}}{2(m-2)G^2\gamma^2}\frac{\Gamma\left(\frac{5}{4}-\frac{m}{8}\right)}{\Gamma\left(\frac{1}{4}\right)}\sum_{k=0}^{\infty}\frac{\Gamma\left(\frac{1}{4}+k\right)\Gamma\left(1+k\right)}{\Gamma\left(\frac{5}{4}-\frac{m}{8}+k\right)k!}B_{-\beta_c c^2}\left(-\frac{m}{8}+k+\frac{3}{4},\frac{m}{8}\right),\label{eq:Inte_phi_c_2}
 \end{equation}
in this scenario, the integration conditions are $\beta_c>0$, $c>0$, and $m<6+8n$, which are also consistent with \eqref{eq:Relation_m_with_n}. The expression $B_z\left(A,C\right)$ represents the incomplete beta function. Finally, beginning with \eqref{eq:psiConju_K_psi}, the expression for $L_2$ becomes
 \begin{multline}
         L_2=\frac{(-\beta_c)^{\frac{1}{8} (m-6)}}{2(m-2)G^2\gamma^2}\frac{\Gamma\left(\frac{5}{4}-\frac{m}{8}\right)}{\Gamma\left(\frac{1}{4}\right)}\sum_{k=0}^{\infty}\frac{\Gamma\left(\frac{1}{4}+k\right)\Gamma\left(1+k\right)}{\Gamma\left(\frac{5}{4}-\frac{m}{8}+k\right)k!}\\
         \times\int_{0}^{\infty}(1+\beta_cc^2)^{\frac{m}{8}-1}c^{-\frac{1}{2}-\frac{m}{4}}B_{-\beta_c c^2}\left(-\frac{m}{8}+k+\frac{3}{4},\frac{m}{8}\right)\ dc.\label{eq:L_2_integrate_for_c}
\end{multline}
It is interesting to note that the integral in \eqref{eq:L_2_integrate_for_c} only converges for $k=0$ within the integration interval fixed by the condition $c>0$ in \eqref{eq:Inte_phi_c_2}. Therefore, the other terms in the summation are omitted, and only the first one survives. With this in mind, the expression for $L_2$ reduces to
\begin{equation}
    L_2=\frac{(-\beta_c)^{\frac{1}{8} (m-6)}}{2(m-2)G^2\gamma^2}\int_{0}^{\infty}(1+\beta_cc^2)^{\frac{m}{8}-1}c^{-\frac{1}{2}-\frac{m}{4}}B_{-\beta_c c^2}\left(-\frac{m}{8}+\frac{3}{4},\frac{m}{8}\right)\ dc.\label{eq:L_2_integrate_for_c_1}
\end{equation}
From here, it's clear that $L_2$ is finite for the allowed values of the quantum number $m$, as established in \eqref{eq:Relation_m_with_n}. This can be verified in Table \ref{tab:Values_of_L2_several_m}, where the value of $L_2$ was calculated for different permitted values of $m$.
\begin{table}[htbp]
\captionsetup{format= hang, justification=raggedright, font=footnotesize}
        \centering
        \begin{tabular}{cc}
                \toprule
                $m$ &                                           $L_2$                                             \\
                \midrule
                -1   &                             $-\frac{0.07}{ \gamma ^2 G^2}$                                       \\
                -3   &                             $-\frac{0.02}{\gamma ^2 G^2}$                       \\
                -5   &                             $-\frac{0.01}{ \gamma ^2 G^2}$                        \\
                -7  &                             $-\frac{0.003}{ \gamma ^2 G^2}$                        \\
                \bottomrule
        \end{tabular}
        \caption{Values of $L_2$ in equation \eqref{eq:L_2_integrate_for_c_1} are presented for fixed $\beta_c=2$, various negative $m$, and $a=1/2$.}
        \label{tab:Values_of_L2_several_m}
\end{table}

Since we established that the denominator of \eqref{eq:Kretsch_Exp_Gen} is finite in the context of the minimal uncertainty approach, the above result shows that \eqref{eq:psiConju_K_psi} converges to zero due to \eqref{eq:L1_integrate_in_b_a_1/2} and \eqref{eq:L_2_integrate_for_c_1}, under the conditions imposed so far. Putting all these together, we conclude that the expectation value of the Kretschmann scalar is vanishing under the conditions
\begin{equation}
        \langle \hat{K}\rangle_m=0\quad \text{for}\quad a=1/2,\,\beta_c>0,\,\beta_b\gamma^2=2,\, m=-2(n+\frac{1}{2}),\, n \in \mathbb{N}.
\end{equation}
This result is completely different from the one obtained in \eqref{eq:Expec_value_K_n_witou_GUP}, in which the expectation value of the Kretschmann scalar diverges. 
Together with the fact that the GUP wave function in \eqref{eq:wave-function-GUP-p_c=00003D0}  remains finite everywhere (while the one in \eqref{eq:Expec_value_K_n_witou_GUP}  diverges at $p_c=0$), the above result implies that the physical singularity has been resolved in this new scheme based on the minimal uncertainty method.

\section{Discussion and conclusion\label{Sec:Conclude}}

Since black holes are assumed to be one of the main arenas of quantum gravity, studying their quantum nature, particularly the quantum state of their interior and the fate of their singularities are a main interest of the quantum gravity community. In this paper we have presented two quantization scheme of the interior and compared their results regarding the fate of the singularity. We have first quantized the interior, written in terms of Ashtekar-Barbero connection variables, using the standard quantization techniques. We found the quantum state of the interior and also computed the expectation value of the Kretschmann scalar for this state. It turns out that both of these quantities diverge at the region where the classical singularity resides. We then, for the first time, present a quantization of the interior in the aforementioned variables using the minimal uncertainty approach. 

\begin{enumerate}
\item We computed the quantum states of the interior and showed that they
(and their modulus squared) are everywhere finite in the interior,
meaning that all the interior states are well-defined and square-integrable.
\item We then computed the expectation value of the Kretschmann scalar on
these states just as we did for the standard quantization scheme.
We showed that contrary to the standard quantization, this expectation
value is finite everywhere within the black hole in the minimal uncertainty
approach, particularly where used to be the classical singularity.
Hence, the finiteness of both the quantum states and the expectation
value of the Kretschmann scalar leads to the conclusion that what
used to be a singularity in the classical theory is now resolved in
the minimal uncertainty quantization scheme.
\item We also found a quantum number $m$ in our model which determines
the convergence of the norm of states, as well as the convergence
and finiteness of the expectation value of the Kretschmann scalar.
This is a new quantum number which was not reported before.
\item Under the condition of convergence of norm of states, we found a minimum
for the square of the radius of the (2-spheres in the) black hole,
$p_{c}$. This minimum radius is universal, meaning that it is independent
of the mass of the black holes and solely depends on one of the minimal
uncertainty scales, $\beta_{c}$, as
\begin{equation}
p_{c}>\frac{\sqrt{\beta_{c}}}{4}.
\end{equation}
\item By demanding square-integrability of the states, particularly at the
location where there used to be a classical singularity, we obtained
an exact relation between the Barbero-Immirzi parameter $\gamma$
of LQG and one of the minimal uncertainty scales, $\beta_{b}$, as
\begin{equation}
\beta_{b}=\frac{2}{\gamma^{2}}.
\end{equation}
\end{enumerate}

This method has the potential to be extended to study the whole spacetime of a black hole, which is a project that we leave for a future work.

\begin{acknowledgements}
S. R. acknowledges the support of the Natural Science and Engineering Research Council of Canada, funding reference No. RGPIN-2021-03644 and No. DGECR-2021-00302. O. O and W. Y thanks to the grant by the University of Guanajuato CIIC 168/2023 "Non-extensive Entropies Independent of Parameters" and CIIC 156/2024 "Generalized uncertainty principle, non-extensive entropies, and general relativity".
 \end{acknowledgements} 

\bibliographystyle{apsrev4-2}
\bibliography{mainbib}

\end{document}